\documentclass{ws-ijmpd}

\usepackage{graphicx}
\usepackage{amsmath}
\usepackage{amssymb}
\usepackage{graphicx}
\usepackage{slashed}
\usepackage{textcomp}

\usepackage{epstopdf}

\usepackage[center]{subfigure}

\begin{document}

%%%%%%%%%%%%%%%%%%%%%%%%%%%%%%%%%%%%%%%%%%%%%%%%%%%%%%%%%%%%%%%
\def\Box{\square}
\def\t#1{\mathrm{#1}}
\def\c#1{\mathcal{#1}}
\def\pd{\partial}
\def\inf{\infty}
\def\e{\epsilon}
\def\->{\rightarrow}
\def\=={\begin{eqnarray*}}
\def\xx{\end{eqnarray*}}
\def\n==#1{\begin{eqnarray} \label{#1}}
\def\nxx{\end{eqnarray}}
\def\half{\frac{1}{2}}
\def\shalf{\textstyle\frac{1}{2}\displaystyle}
\def\sfrac#1#2{\textstyle\frac{#1}{#2}\displaystyle}

\makeatletter
\def\eqnarray{ \stepcounter{equation} \let\@currentlabel=\theequation
 \global\@eqnswtrue
 \global\@eqcnt\z@
 \tabskip\@centering
 \let\\=\@eqncr
 $$\halign to \displaywidth\bgroup\@eqnsel\hskip\@centering
 $\displaystyle\tabskip\z@{##}$&\global\@eqcnt\@ne
 \hfil$\displaystyle{{}##{}}$\hfil
 &\global\@eqcnt\tw@$\displaystyle\tabskip\z@{##}$\hfil
 \tabskip\@centering&\llap{##}\tabskip\z@\cr}
\makeatother

%%% new array
\makeatletter
\def\@arrayacol{\edef\@preamble{\@preamble \hskip .5\arraycolsep}}
\def\array{\let\@acol\@arrayacol \let\@classz\@arrayclassz
\let\@classiv\@arrayclassiv \let\\\@arraycr\def\@halignto{}\@tabarray}
\makeatother
%%%

%%% new arraystretch
\renewcommand{\arraystretch}{1.6}
%%%

%%% subequations
\makeatletter
\newcounter{subeqncnt}
\def\thesubeqncnt{\alph{subeqncnt}}
\def\subequations{\begingroup%
   \stepcounter{equation}\edef\@tempa{\theequation}%
   \let\c@equation\c@subeqncnt\c@subeqncnt\z@
   \edef\theequation{\@tempa\noexpand\thesubeqncnt}}
\let\endsubequations\endgroup
\makeatother
%%%%%

%%% hangcaption
\newcommand{\captionfonts}{\small}
\makeatletter % Allow the use of @ in command names
\long\def\@makecaption#1#2{%
\vskip\abovecaptionskip
\sbox\@tempboxa{{\captionfonts #1: #2}}%
\ifdim \wd\@tempboxa >\hsize {\captionfonts #1: #2\par} \else
\hbox to\hsize{\hfil\box\@tempboxa\hfil}%
\fi \vskip\belowcaptionskip}
\makeatother % Cancel the effect of \makeatletter
%%%%%

%%% newcommand
\newcommand{\del}{\partial}
\newcommand{\ep}{\varepsilon}
\newcommand{\tr}{{\rm Tr}\,}
\newcommand{\sdet}{{\rm sdet}\,}
\newcommand{\dd}{{\rm d}}

\newcommand{\mbar}[1]{\overline {#1} \hskip 1pt{}}
\newcommand{\bra}[1]{ \langle #1 | }
\newcommand{\ket}[1]{ | #1 \rangle }
\newcommand{\vac}{ \bra{{\rm vac}} }
\newcommand{\cuum}{ \ket{{\rm vac}} }
\newcommand{\expect}[1]{ \langle \, #1 \, \rangle }

\newcommand{\define}{ \stackrel{\rm def}{\equiv} }
\newcommand{\intdzeta}{ { \int\limits_{- i \infty}^{+ i \infty}
                        \frac{d \zeta}{2 \pi i} } }
\newcommand{\intz}{ { \oint\!\frac{dz}{2 \pi i} } }
\newcommand{\prtz}{ { \frac{\prt}{\prt z} } }
\newcommand{\ZZ}{{\bf Z}}
\newcommand{\const}{{\rm const.}}

\def\imo{i}
 %%%%%%%%%%%%%%%%%%%%%%%%%%%%%%%%%%%%%%%%%%%%%%%%%%%%%%%%%%%%%%%

\title{Causality Constraint on Noncritical Einstein-Weyl Gravity}

\author{Fu-Wen Shu$^{1,2}$}

\address{$^1$Center for Relativistic Astrophysics and High Energy
Physics,
Department of Physics,\\ Nanchang University, Nanchang, 330031, China\\
 $^2$College
of Mathematics and Physics,Chongqing University of
Posts and Telecommunications, Chongqing, 400065, China\\
fwenshu@gmail.com}

\author{Yungui Gong}

\address{School of Physics, Huazhong University of Science and
Technology, Wuhan, 430074, China\\
yggong@hust.edu.cn}

\maketitle

\begin{abstract}
We explore, in the context of AdS/CFT correspondence, the causality
constraints on the Noncritical Einstein-Weyl (NEW) gravity model in
five dimensions. The scalar and shear channels are considered as
small metric perturbations around an AdS black brane background. Our
results show that causality analysis on the propagation of these two
channels imposes a new bound on the coupling of the Weyl-squared
terms in the NEW gravity.  This new bound imposes more stringent
restrictions than those of the tachyon-free condition, improving
predictive power of the theory.
\end{abstract}


% \tableofcontents

%%%%%
\section{Introduction}
\setcounter{equation}{0} \setcounter{footnote}{0}

Recently much attention has been paid to topologically massive
gravity\cite{djt} in three dimensions. The model contains a massive
spin-$2$ mode in addition to the usual massless graviton. The
situations are changed after a special Chern-Simons coupling has
been chosen. The extra massive spin-$2$ mode becomes massless, and
the model reduces to the standard Einstein gravity after truncating
out the logarithmic fall-off modes by imposing an appropriate
boundary condition \cite{lss}. Therefore, the model is a toy model
for a quantum theory of gravity due to its possibly well-controlled
UV behaviour.

A recent generalization of this ``critical'' theory to four
\cite{lu1} and higher dimensions \cite{lu2} results in a model known
as ``critical gravity''. This is a theory of gravity which includes
Einstein-Hilbert term, cosmological constant and Weyl-squared terms
in its action. In this model a special value (critical point) of the
coupling of the Weyl-squared terms is chosen so that the massive
spin-2 mode becomes a logarithmic mode. The new model presents many
remarkable properties: (i) The model is perturbatively
renormalizable since the action includes curvature-squared
corrections as studied in detail in \cite{stelle}; (ii) Due to the
special coupling of Weyl-squared terms, the model contains a
massless graviton and a logarithmic mode, which distinguishes it
from most of models with higher derivative terms; (iii) The massless
graviton has zero on-shell energy, making the theory trivial. More
recent discussion on this model can be found in a lot of literature
\cite{Pang:2011cs}${}^\sim$\cite{liu}.

Recent progress\cite{Lu:2011ks,lu3} shows that by abounding the
severe constraint imposed on the value of the coupling, one can
generalize critical gravity to the so-called Noncritical
Einstein-Weyl (NEW) gravity. This is a wider class of Weyl-squared
extensions to Einstein gravity with cosmological constant. The value
of the Weyl-squared coupling in this model is no longer fixed at the
critical point, instead it relaxes to a certain range with which the
massive mode in the AdS background is still tachyon-free. While this
mode is still ghostlike, it can be truncated by imposing appropriate
boundary conditions since this massive mode falls off more slowly
than the massless graviton. Then only the massless graviton survives
in this theory and this mode, different from critical gravity, has
the positive excitation energy.

Inspired by the remarkable observation in \cite{Maldacena:2011mk},
authors of \cite{HJJY,HJJY2} showed that the theory may also have
the same physics as Einstein gravity on AdS background. They found
that the effective Lagrangian of NEW gravity is identical with the
one of Einstein gravity up to the rescaling of Newton's constant.
From the perspectives of the AdS/CFT correspondence \cite{ads/cft},
the classical equivalence between NEW gravity and Einstein gravity
tells us that the corresponding dual conformal field theories(CFT)
are equivalent classically in the large $N$ limit. Holographic
description of this theory has been investigated partially in
\cite{HJJY,HJJY2}.

One important progress for the application of the AdS/CFT
correspondence is that inclusion of higher-order curvature terms in
the Einstein-Hilbert action may possibly introduce the superluminal
propagation of boundary perturbations \cite{kp}${}^\sim$\cite{sin2}.
In order to preserve causality at the boundary, constraints have to
be imposed on the couplings of the higher-order terms. It was first
realized by Horman and Maldacena in \cite{hofman} and then by Buchel
and Myers in \cite{BM} that the causality constraints are closely
related to the constraints\footnote{This constraints come from the
requirement that the energy measured in calorimeters of a collider
physics experiment should be positive\cite{hofman}.} imposed on the
ratio of the central charges, $a/c$, in the dual CFT.

In this paper we would like to consider the causality constraints,
in the context of the AdS/CFT correspondence, on the couplings of
the Weyl-squared terms. Our calculation are performed in the gravity
side, by studying their metric fluctuations around an AdS black
brane background. Our results show that to preserve the causality at
the boundary, new bounds on the Weyl-squared couplings are
introduced.

The rest of the paper is organized as follows. In the next section
we give a brief review on the Einstein-Weyl gravity, and the bounds
on the Weyl-squared coupling are given by requiring that the massive
mode is tachyon-free. In section three we find an approximate black
brane solution to the NEW gravity model. In section four we consider
a small metric fluctuations around the black brane, and a set of
linearized wave equations are obtained for tensor-type and
vector-type perturbations. In section five causality constraints on
the couplings are discussed for both the scalar channel and shear
channel. Conclusions and discussions are presented in the last
section.

\section{Noncritical Einstein-Weyl Gravity: A Brief Review}
\noindent In this section we would like to give a brief review on
NEW gravity. The starting point is a gravitational action containing
higher curvature correction terms. We generally pay much attention
to the quadratic corrections only. The most general quadratic
gravity model with a cosmological constant is given by
\begin{equation} I=\int
d^{D}x\,\sqrt{-g}\left[\frac{1}{\kappa}\left(R-2\Lambda\right)+\alpha
R^{2}+\beta R^{\mu\nu}R_{\mu\nu}+\gamma\mathcal{L}_{GB}\right],
\label{eq:Quadratic_action}
\end{equation}
where $\kappa=16\pi G$ and $\mathcal{L}_{GB}$ is the Gauss-Bonnet
term
$$
\mathcal{L}_{GB}=R^{\mu\nu\rho\sigma}R_{\mu\nu\rho\sigma}-4R^{\mu\nu}R_{\mu\nu}+R^{2}.
$$
This term can be generated from the low energy effective theory of
heterotic and bosonic string theory, and it is a topological
invariant in four dimensions.

It is straightforward to show that the equations of motion that
follow from the action (\ref{eq:Quadratic_action}) are
\begin{eqnarray}
\label{eom} R_{\mu\nu}-\frac{1}{2}g_{\mu\nu}R+g_{\mu\nu}\Lambda
=\kappa T^{\rm eff}_{\mu\nu},
\end{eqnarray}
where
\begin{eqnarray}\label{effT}
\nonumber T_{\mu\nu}^{\rm eff} &=& \frac{\alpha}2 \left(g_{\mu\nu}
R^2 - 4RR_{\mu\nu} + 4\nabla_\nu\nabla_\mu R - 4g_{\mu\nu}\Box R
\right) + \nonumber \\ \nonumber &+& \frac{\beta}2 \left(g_{\mu\nu}
R_{\rho\sigma} R^{\rho\sigma} + 4\nabla_\alpha\nabla_{(\nu}
{R_{\mu)}} ^\alpha - 2\Box R_{\mu\nu} -
g_{\mu\nu} \Box R - 4R^\alpha_\mu R_{\alpha\nu} \right) + \\
&+& \frac{\gamma}2 \left(g_{\mu\nu}\mathcal{L}_{GB} -
4R_{\mu\alpha\beta\gamma}{R_\nu}^{\alpha\beta\gamma} - 4RR_{\mu\nu}+
8R_\mu^\alpha R_{\alpha\nu}
+ 8R^{\alpha\beta}R_{\mu\alpha\nu\beta} \right),\nonumber\\
\end{eqnarray}
with $\nabla$ the covariant derivative and $\Box\equiv \nabla^2$ the
d 'Alembert operator. For Eintein-Weyl gravity, we have
$$\alpha=-\frac{D(D-3)\gamma}{(D-1)(D-2)},\ \ \ \ \ \
\beta=\frac{4(D-3)\gamma}{D-2},
$$
and the action reduces to
\begin{equation}\label{weyl action}
S=\int d^D x\sqrt{-g}\left[\frac1{\kappa}(R-2\Lambda)+\gamma
C_{\mu\nu\rho\sigma}C^{\mu\nu\rho\sigma}\right],
\end{equation}
where $C_{\mu\nu\rho\sigma}$ is the Weyl conformal tensor and
satisfies
\begin{equation}
 C_{\mu\nu\rho\sigma}C^{\mu\nu\rho\sigma}=R_{\mu\nu\rho\sigma}R^{\mu\nu\rho\sigma}-\frac{4}{D-2}R_{\mu\nu}R^{\mu\nu}+\frac{2}{(D-1)(D-2)}R^2.
\end{equation}
The equation of motion following this action then becomes
\begin{eqnarray}
\label{eom_weyl} R_{\mu\nu}-\frac{1}{2}g_{\mu\nu}R+g_{\mu\nu}\Lambda
=\kappa T^{\rm Weyl}_{\mu\nu},
\end{eqnarray}
where
\begin{eqnarray}\label{weylT}
\nonumber T_{\mu\nu}^{\rm Weyl} &=&\gamma\left\{
-\frac1{(D-1)(D-2)}R(4R_{\mu\nu}-g_{\mu\nu}
R)-\frac2{D-2}[g_{\mu\nu} R_{\rho\sigma} R^{\rho\sigma}
-2(D-2)R^\rho_\mu R_{\rho\nu}]\right.
\nonumber\\
&&\left.+\frac{2(D-3)}{(D-1)(D-2)}[g_{\mu\nu} \Box R +
(D-2)\nabla_{\mu}\nabla_{\nu}R]- \frac{4(D-3)}{D-2} \Box R_{\mu\nu}
\right.\nonumber\\ &&
\left.+\frac12g_{\mu\nu}R_{\rho\sigma\lambda\kappa}R^{\rho\sigma\lambda\kappa}
-\frac{4(D-4)}{D-2}R^{\rho\sigma}R_{\mu\rho\nu\sigma}
-2R_{\mu\rho\sigma\gamma}R_{\nu}{}^{\rho\sigma\gamma}\right\}.
\end{eqnarray}

One of the biggest issues with Einstein-Weyl gravity theory, just
like most of the higher derivative theories \footnote{Ghost-free
conditions for higher derivative theories are discussed recently in
\cite{bgkm}.}, is the presence of ghosts, which point to
instabilities of the quantum version of the theory. To see this
clearly, we note that the Einstein-Weyl theory admits AdS spacetime
as the vacuum solution $\bar{g}_{\mu\nu}$ of the field equations,
and we have
\begin{equation}
\bar{R}=\frac{2D}{D-2}\Lambda,\ \ \
\bar{R}_{\mu\nu}=\frac{2\Lambda}{D-2} \bar{g}_{\mu\nu},\ \ \
\bar{R}_{\mu\nu\rho\sigma}=\frac{2\Lambda}{(D-1)(D-2)}(\bar{g}_{\mu\rho}\bar{g}_{\nu\sigma}-\bar{g}_{\mu\sigma}\bar{g}_{\nu\rho}).\end{equation}
By considering the metric fluctuations around this AdS vacuum
$g_{\mu\nu}=\bar{g}_{\mu\nu}+h_{\mu\nu}$, we obtained linearized
equation of motion
\begin{equation}
\left(\bar{\Box}-\frac{4\Lambda}{\left(D-1\right)\left(D-2\right)}-M^{2}\right)\left(\bar{\Box}-\frac{4\Lambda}{\left(D-1\right)\left(D-2\right)}\right)h_{\mu\nu}=0,\label{4eom}\end{equation}
 where $\bar{\Box}=\bar{\nabla}^2$ with $\bar{\nabla}$ the covariant derivative with respect to $\bar{g}_{\mu\nu}$ and $M^2$ in (\ref{4eom}) is
defined as
\begin{equation}
M^{2}\equiv
-\frac{D-2}{4(D-3)}\left(\frac1{\kappa\gamma}-\frac{8(D-3)\Lambda}{(D-1)(D-2)}\right)=m^2-\frac{D-2}{L^2},\label{eq:Mass_of_spin-2_excitation}\end{equation}
where $L$ is the radius of the AdS space and
$$
m^2\equiv -\frac{D-2}{4(D-3)\kappa\gamma}.
$$

Note that in deriving (\ref{4eom}) the transverse and traceless
gauge, which is consistent with the equation of motion, has been
used
\begin{equation}
\bar\nabla^{\mu}h_{\mu\nu}=0,\qquad
h=0.\label{TTgauge}\end{equation} As a consequence, the fourth order
differential equation (\ref{4eom}) of the Einstein-Weyl gravity
consists of a massless mode $h_{\mu\nu}^{\left(m\right)}$ and a
massive mode $h_{\mu\nu}^{\left(M\right)}$ which satisfy,
respectively, the wave equations
\begin{eqnarray}
\left(\bar{\square}-\frac{4\Lambda}{\left(D-1\right)\left(D-2\right)}\right)h_{\mu\nu}^{\left(m\right)}&=&0,\\
\left(\bar{\square}-\frac{4\Lambda}{\left(D-1\right)\left(D-2\right)}-M^{2}\right)h_{\mu\nu}^{\left(M\right)}&=&0.\label{mmodes}\end{eqnarray}
In a special case where $M^{2}=0$, i.e.
\begin{equation}m^2=\frac{D-2}{L^2}\ \ \  \ \mbox{or} \ \ \ \gamma=-\frac{L^2}{4(D-3)\kappa}, \end{equation}
the two linearized equations of motion degenerate. This case is
generally called critical gravity\cite{lu1,lu2}. The second mode
solution in this case is no longer massive, instead, it becomes a
logarithmic mode in the sense that this mode
$h_{\mu\nu}^{\left(L\right)}$ satisfies a quadratic differential
equation,
\begin{equation}
\left(\bar{\square}-\frac{4\Lambda}{\left(D-1\right)\left(D-2\right)}\right)^2h_{\mu\nu}^{\left(L\right)}=0.
\end{equation}
It is generally believed that the logarithmic mode can be truncated
by imposing a suitable boundary condition at the conformal boundary
due to the fact that the logarithmic mode falls off more slowly than
massless mode as it approaches the boundary. Recent
progress\cite{lu1,liu} shows, however, that Einstein-Weyl gravity at
the critical point seems to be trivial since the massless mode has
zero energy of excitations. As a consequence, particular attention
has been paid to the case with $-\frac{(D-3)^2}{4L^2}\leq
m^2<\frac{D-2}{L^2}$ where the theory is still tachyon-free since it
is above the Breitenlohner-Freedman bound\cite{BF}\footnote{This is
a bound above which the negative mass for a bulk field in AdS
background will not lead to any instabilities. To obtained the
expression for this bound, one can first rewrite Eq.(\ref{mmodes})
as a standard Schr\"{o}dinger equation by making a Fourier
transformation. Then suitable boundary condition should be imposed
so that we have a normalizable solution to the AdS wave equation.
Once we do so, we find that for any $\alpha=M^2-(1-(D-1)^2)/4>-1/4$
we cannot find a normalizable negative-energy state on which the
Schrodinger operator is Hermitian. Therefore we obtain the bound
\cite{BF}
$$M^2\geq-\frac{(D-1)^2}{4L^2}.$$
Equivalently, we have the bound for $m^2$
$$m^2\geq-\frac{(D-3)^2}{4L^2}.$$}. In this case, we notice that the
massive mode falls off less rapidly at the boundary than massless
mode since $M^2<0$. Therefore, one can truncate the massive mode by
imposing suitable boundary condition as before and the massless mode
with non-zero excitation energy survives only. In what follows, we
mainly focus on the Einstein-Weyl gravity with this coupling, i.e.,
\begin{equation}\label{tachyonfree}
-\frac{L^2}{4(D-3)\kappa}<\gamma\leq \frac{(D-2)L^2}{(D-3)^3\kappa}.
\end{equation}
The Einstein-Weyl gravity with this coupling is often referred to as
NEW gravity.

\section{AdS Black Brane in NEW Gravity}
\noindent In this
section we would like to find an AdS black brane solution to the NEW
gravity. Owing to the fact that the field equations of Einstein-Weyl
gravity are a set of fourth order differential equations, it is
generally quite hard to obtain their exact solutions. There are some
exact solutions of pure conformal gravity (i.e., the limited case of
Einstein-Weyl gravity when $\gamma$ goes to infinity) in four
dimensions in the literatures, say \cite{klemm}. However, it is very
difficult to obtain an exact solution to NEW gravity with the
presence of Ricci scalar and cosmological constant in any dimension.
For simplicity, we find our AdS black brane solution approximately.

At the beginning, we note that for the following Einstein-Hilbert
action with negative cosmological constant
\begin{equation}
I=\frac{1}{\kappa}\int
d^{D}x\,\sqrt{-g}\left(R-2\Lambda\right),\label{EH_action}
\end{equation}
there is an AdS black brane solution in Poincar\'{e} coordinates
\n=={metric-unperturbed} ds^2 = \frac{1}{z^2}\left(\frac{-f(z) dt^2
+ d\vec x^2}{L^2} + \frac{L^2 dz^2}{f(z)}\right), \nxx where
\n=={ffunc} f(z)=1-\left(\frac{z}{z_0}\right)^{D-1}, \nxx with $z_0$
the horizon of the black brane.

The corrected black brane metric of the NEW gravity can be obtained
by noting that all terms in $T_{\mu\nu}^{Weyl}$ are higher order
corrections as shown in (\ref{weylT}). Hence one can calculate
$T_{\mu\nu}^{Weyl}$ by substituting the unperturbed metric
(\ref{metric-unperturbed}). A perturbatively approximate solution of
the field equation (\ref{eom_weyl}) calculated in this way is given
by (for details, see Appendix B in \cite{kp})
\n=={metric-z} ds^2 = \frac{1}{z^2}\left(\frac{-N^2f(z) dt^2 + d\vec
x^2}{L^2} + \frac{L^2 dz^2}{f(z)}\right) \nxx where $N^2$ is a
constant whose value will be determined later and \n=={f(z)-app}
f(z) = 1 - \left(\frac{z}{z_0}\right)^{D-1}  + \lambda
\left(\frac{z}{z_0}\right)^{2(D-1)} \nxx  with $\lambda$ a
dimensionless constant \== \lambda =
\frac{(D-3)(D-4)}{L^2}\kappa\gamma. \xx If we change the variable
$z$ to $u$ as \== u = \left(\frac{z}{z_0}\right)^\frac{D-1}{2}, \xx
we obtain the following metric \n=={metric} ds^2 =
\frac{-N^2f(u)\,dt^2 + d\vec x^2}{L^2z_0^2\,u^\frac{4}{D-1}} +
\frac{4L^2}{(D-1)^2}\frac{du^2}{u^2 f(u)} \,, \nxx
where  $f(u)$ now becomes \n=={f(u)} f(u) = 1 - u^2  + \lambda u^4.
\nxx

In these coordinates, $u = 0$ is the boundary of the AdS space, and
the horizon of the black brane is at \n=={horizon} u_H \simeq 1 +
\frac{ \lambda}{2} \nxx

Taking the limit $\lambda\rightarrow 0$, the solution corresponds to
Schwarzschild-AdS spacetime. %The causality analysis in this
%background has been well investigated in \cite{shenker}.

To determine the value of the constant $N^2$ in the metric
(\ref{metric}), we notice that the space-time geometry of the
background at the boundary would reduce to flat Minkowski metric
conformally, i.e.\ $\dd s^2\propto -c^2\dd t^2+\dd\vec{x}^2$. On the
boundary $r\rightarrow\infty$, we have
$$
N^2f(u\rightarrow 0) \rightarrow 1,
$$
so that $N^2$ is found to be
\begin{equation}
N^2=1,
\end{equation}
by specifying the boundary speed of light to be unity $c=1$.
Therefore, the metric of the black brane in (\ref{metric}) is now of
the form \n=={metric-u} ds^2 = \frac{-f(u)\,dt^2 + d\vec
x^2}{L^2z_0^2\,u^\frac{4}{D-1}} +
\frac{4L^2}{(D-1)^2}\frac{du^2}{u^2 f(u)} \,. \nxx This is the AdS
black brane solution which will be mainly considered in the
following.

\section{Linearized Wave Equation in the AdS Black Brane Background}
To see the propagation of the transverse graviton, in this section
we shall consider small metric fluctuations $h_{\mu\nu}$ around the
AdS black brane background (\ref{metric-u})
\begin{eqnarray}
g_{\mu\nu} &\equiv& g^{(0)}_{\mu\nu}+h_{\mu\nu},
\end{eqnarray}
where the background metric $g^{(0)}_{\mu\nu}$ is given in
(\ref{metric-u}). To the first order of the metric perturbation, one
can define an inverse metric as
\begin{equation}
g^{\mu\nu}=g^{(0)\mu\nu}-h^{\mu\nu} + {\cal O}(h^2),
\end{equation}
and the indices are raised and lowered by using these background
metric $g_{\mu\nu}^{(0)}$ and $g^{(0)\mu\nu}$. We also denote a
trace part of the metric and a field strength for the perturbative
parts as $h\equiv g^{(0)\mu\nu} h_{\mu\nu}$.

Now we would like to think of a linearized theory of the symmetric
tensor field $h_{\mu\nu}$ propagating in the   AdS black brane
background. To the first order of $h_{\mu\nu}$, the Einstein
equation (\ref{eom_weyl}) can be written as
\begin{equation}
R^{(1)}_{\mu\nu} -\frac{1}{2}g^{(0)}_{\mu\nu}R^{(1)}
-\frac{1}{2}h_{\mu\nu}R^{(0)} +h_{\mu\nu}\Lambda =\kappa
T^{(1)}_{\mu\nu} \label{eom_01}
\end{equation}
In the expression above,
%$R_{mn}^{(0)}(x)$,
the scalar curvature $R^{(0)}$
%and $T^{(0)}_{mn}(x)$ are
is constructed by using the background metric $g_{\mu\nu}^{(0)}$ and
the following tensors are newly defined:
\begin{eqnarray}
R^{(1)}_{\mu\nu} &=& \frac{1}{2} \Big(\nabla_\rho\nabla_\mu
h_\nu{}^\rho +\nabla_\rho\nabla_\nu h_\mu{}^\rho-\nabla^2h_{\mu\nu}
-\nabla_\mu\nabla_\nu h \Big),
\\
R^{(1)} &=&\nabla_{\mu}\nabla_{\nu}h^{\mu\nu}-\nabla^2h
-h^{\mu\nu}R_{\mu\nu}^{(0)},
\end{eqnarray}
where $\nabla$ is the covariant derivative with respect to the
unperturbative AdS background  $g_{\mu\nu}^{(0)}$. In the right hand
side of (\ref{eom_01}), $T^{(1)}_{\mu\nu}$ denotes the linearized
terms of $T^{\rm Weyl}_{\mu\nu}$ in (\ref{eom_weyl}). The explicit
expression for $T^{(1)}_{\mu\nu}$ is given in Appendix
(\ref{expr_T1}).

It is obvious that the linearized wave equation (\ref{eom_01}) in
general includes derivative terms with order higher than two which
may make it difficult to study the propagating graviton. Our
strategy of finding a second order differential equation is as
follows. We first note that higher order (greater than 2)
derivatives of the metric fluctuations come from  sources $\Box
R^{(1)}$, $\nabla_\mu\nabla_\nu R^{(1)}$ and $\Box R^{(1)}_{\mu\nu}$
in (\ref{higher1})-(\ref{higher2}) in appendix (\ref{expr_T1}). It
follows from equation (\ref{eom_01}) that one can replace $R^{(1)}$
and $R^{(1)}_{\mu\nu}$ approximately by their leading order values
by taking $\lambda\rightarrow 0$ (or $\gamma\rightarrow 0$),
\begin{eqnarray}
R^{(1)} &\simeq& \frac{R^{(0)}-2\Lambda}{2-D}h\\
R^{(1)}_{\mu\nu}&\simeq&
\frac12\left(\frac{R^{(0)}-2\Lambda}{2-D}\right)g^{(0)}_{\mu\nu}h+\frac12
R^{(0)}h_{\mu\nu}-\Lambda h_{\mu\nu}.
\end{eqnarray}
In this way all derivatives become the second order
\begin{eqnarray}
\Box R^{(1)}&=&\frac{1}{2-D}\left(\Box R^{(0)} h+2\nabla_\mu
R^{(0)}\nabla^\mu h+
(R^{(0)}-2\Lambda)\Box h\right)\\
\nabla_\mu\nabla_\nu
R^{(1)}&=&\frac{1}{2-D}\left(\nabla_\mu\nabla_\nu R^{(0)}
h+\nabla_\mu R^{(0)}\nabla_\nu h+\nabla_\nu R^{(0)}\nabla_\mu h+
(R^{(0)}-2\Lambda)\nabla_\mu\nabla_\nu h\right)\\
\nonumber \Box
R^{(1)}_{\mu\nu}&=&\frac{g^{(0)}_{\mu\nu}}{4-2D}\left(\Box R^{(0)}
 h+2\nabla_\mu
R^{(0)}\nabla^\mu h+(R^{(0)}-2\Lambda)\Box h\right)+\\
 &&+\frac12 \left(\Box
R^{(0)}h_{\mu\nu}+2\nabla_\mu R^{(0)}\nabla^\mu h_{\mu\nu}+
R^{(0)}\Box h_{\mu\nu}\right)-\Lambda \Box h_{\mu\nu}.
\end{eqnarray}
And the linearized equation of motion can be obtained by
substituting the perturbative metric.

Generally speaking, there are three types of metric perturbations:
scalar perturbation, vector perturbation and tensor perturbation. In
the present paper, we shall work in the $h_{u\mu}=0$ gauge and use
the Fourier decomposition
\begin{eqnarray*}
h_{\mu\nu}(t, x_3, u) &=& \int\!\frac{\dd \omega\dd k}{4\pi^2}
e^{-i\omega t+ikx_3}h_{\mu\nu}(u),
\end{eqnarray*}
where we choose the momenta to be along the $x_3$-direction, and for
the sake of simplicity we have fixed $D=5$ as an example
\footnote{Hereafter we will pay our attention to the case with
$D=5$. Higher dimensions will be investigated in the future.}.

In this gauge, one can categorize the metric perturbations to the
following three types according to the $O(2)$ rotation in the $(x,
y)$-plane ~\cite{pss}:
\begin{itemize}
\item tensor type (scalar channel): \
$h_{x_1x_2}\ne0$, \ $h_{x_1x_1}=-h_{x_2x_2}\ne0$, \
${\mbox{(others)}}=0$
\item vector type (shear channel): \
$h_{x_1t}\ne0$, \ $h_{x_1x_3}\ne0$, \ ${\mbox{(others)}}=0$
\\
\hspace*{27mm} $\Big($equivalently, $h_{x_2t}\ne0$, \
$h_{x_2x_3}\ne0$, \ ${\mbox{(others)}}=0$$\Big)$
\item scalar type (sound channel): \
$h_{tz}\ne0$, $h_{tt}\ne0$, $h_{x_1x_1}=h_{x_2x_2}\ne0$, and
$h_{x_3x_3}\ne0$, \ $\mbox{(others)}=0$
\end{itemize}
In what follows we mainly focus on the first two kinds of
perturbations, leaving the scalar type for future work\footnote{The
scalar type perturbation includes four independent fluctuation
variables, governed by four coupled wave equations. The complexity
makes it much harder to consider in the present context.}.
\subsection{Tensor-type perturbation}
In this subsection we pay our attention to the tensor-type
perturbation. A nontrivial equation of motion in (\ref{eom_01}) is
coming from ($x_1, x_2$) component. In particular, we mainly focus
on the small metric fluctuation $h^{x_1}_{x_2}(t,x_3,u)\equiv\phi(t,
x_3, u)$ around the AdS black brane background(\ref{metric-u}) of
the form
\begin{equation}\label{metric_pert}
\dd s^2 =\frac{-f(u)\dd t^2+\dd{\vec{x}}^2+2\phi(t,x_3,u)\dd x_1\dd
x_2} {L^2z_0^2u}+\frac{L^2 \dd u^2}{4u^2 f(u)}.
\end{equation}
%
%By considering the spin under the $O(2)$ rotation in $(x_1, x_2)$-plane,
%gauge perturbations would be decoupled within this tensor type perturbation.
According to the AdS/CFT correspondence, the fluctuation field
$\phi(t,z,u)$ corresponds to the component $T_{12}$ of the
energy-momentum tensor of the boundary theory.

After using Fourier decomposition
$$
\phi(t, x_3, u) = \!\int\!\frac{\dd \omega\dd k}{4\pi^{2}}
\mbox{e}^{-i\omega t+ikz}\phi(u),
$$
we can obtain the following linearized equation of motion for
$\phi(u)$ from the equation (\ref{eom_01}) for $D=5$:
\begin{eqnarray}
0 &=& \phi''(u) +\left(\frac{g_T'(u)+32\kappa\gamma
u^{3/2}f\left(u^{-1/2}f''(u)\right)'}{g_T(u)}\right)\phi'(u)
\nonumber
\\
&& +\frac{\bar{\omega}^2}{uf^2(u)}\phi(u)-\frac{\bar{k}^2C_k(u)}{u^2
g_T(u)}\phi(u)+I(u)\phi(u), \label{maineq}
\end{eqnarray}
where
$$
g_T(u)\equiv 3L^2u^{-1}f(u)\left[1-\frac{\lambda}6 \left(96-136f(u)+
74uf'(u)-20u^2f''(u)\right)\right],
$$
$$
C_k(u)=3L^2\left[1-\frac{\lambda}6
\left(96-136f(u)+106uf'(u)-28u^2f''(u)\right)\right],
$$
$$
\bar{\omega}\equiv \frac{z_0L^2}{2}\omega, \qquad
\bar{k}\equiv\frac{z_0L^2}{2}k,
$$
and $I(u)$ is a term independent of $\omega$, $k$, $\phi(u)$ and its
derivatives. The explicit expression of $I(u)$ is not important in
our following discussion.

\subsection{Vector-type perturbation}
In this case, nonzero small metric fluctuations are \
$h_{x_1t}(x)\ne 0$, $h_{x_1x_3}(x)\ne 0$. For convenience, we set
$h^{x_1}_t(u)=g^{(0)x_1x_1}h_{x_1t}(u)=\phi(u)$,
$h^{x_1}_{x_3}(u)=g^{(0)x_1x_1}h_{x_1x_3}(u)=\psi(u)$. Equivalently,
the perturbative metric is given by
\begin{equation}\label{metric_pert}
\dd s^2 =\frac{-f(u)\dd t^2+\dd{\vec{x}}^2+2\phi(t,x_3,u)\dd x_1\dd
t+2\psi(t,x_3,u)\dd x_1\dd x_3} {L^2z_0^2u}+\frac{L^2 \dd u^2}{4u^2
f(u)}.
\end{equation}
%
%By considering the spin under the $O(2)$ rotation in $(x_1, x_2)$-plane,
%gauge perturbations would be decoupled within this tensor type perturbation.
Nontrivial equations in the equation of motion (\ref{eom_01}) appear
from $(t, x)$, $(u, x)$ and $(x, z)$ components. It is not difficult
to show that only two of these three linearized equations are
independent. Without loss of generality in the present paper we
choose equations coming from the $(t, x)$ and $(u, x)$ components.
After making Fourier decompositions
\begin{eqnarray*}
\phi(t, x_3, u) &=& \!\int\!\frac{\dd \omega\dd k}{4\pi^{2}}
\mbox{e}^{-i\omega t+ikx_3}\phi(u),\\
\psi(t, x_3, u) &=& \!\int\!\frac{\dd \omega\dd k}{4\pi^{2}}
\mbox{e}^{-i\omega t+ikx_3}\psi(u),
\end{eqnarray*}
they are in general of the following forms:
\begin{eqnarray}
0 &=& \phi''(u) +A_1(u)\phi'(u)+
\Big(A_2(u)\omega^2+A_3(u)k^2+A_4(u)\Big)\phi(u)+A_5(u)k\omega\psi(u)\label{eom_xt},
\\
\nonumber 0 &=& \omega\phi'(u)+k f(u) \psi'(u) +\frac{\lambda L^2}2
\Big[B_1(u)\omega\phi'(u)+ B_2(u)k
\psi'(u)\\
&&+B_3(u)\omega\phi(u)+B_4(u)k\psi(u)\Big], \label{eom_xu}
\end{eqnarray}
where $A_i(u)$ and $B_i(u)$ are some functions of $u$. Explicit
expressions for $A_2(u)$, $A_3(u)$, $A_5(u)$, $B_1(u)$ and $B_2(u)$,
are given in Appendix (\ref{expr for A}). Expressions for $A_1(u)$,
$A_4(u)$, $B_3(u)$ and $B_4(u)$ are irrelevant to our discussion
below where we only focus on the large momenta limit.

Now we can combine the above set of equations into one third order
differential equation for $\phi(u)$. This can be achieved by noting
that $\psi(u)$ in (\ref{eom_xu}) has a perturbative solution in
terms of $\phi(u)$ as we expand it to the order of
$\mathcal{O}(\lambda^2)$,
\begin{equation}\label{phi_psi}
\psi'(u)=-\frac{\omega}{kf(u)}\phi'(u)+\frac{\lambda
L^2}2\left[\left(-\frac{\omega B_1(u)}{kf(u)}+\frac{\omega B_2(u)}{k
f(u)^2}\right)\phi'(u)+\cdots\right]+\mathcal{O}(\lambda^2),
\end{equation}
where ``$\cdots$" denotes terms that are irrelevant to our final
results with the large momenta limit. Differentiating (\ref{eom_xt})
with respect to $u$ and substituting (\ref{phi_psi}) into it we
obtain
\begin{eqnarray}\label{eom_vector}
0 &=& \phi'''(u) +\tilde{A}_2(u)\omega^2\phi'(u)+A_3(u)k^2
\phi'(u)+\cdots +\mathcal{O}(\lambda^2),
\end{eqnarray}
where again ``$\cdots$" are irrelevant terms to our results and
\begin{equation}\label{A2}
\tilde{A}_2(u)=A_2(u)-\frac{A_5(u)}{f(u)}+\frac{3\lambda
z_0^2L^8\left(B_2(u)-B_1(u)f(u)\right)}{8uf(u)^3g_V(u)},
\end{equation}
with $g_V(u)$ given in (\ref{gv}).

\section{Causality Violation}

It was shown that the causality could be violated if higher
derivative terms are considered~\cite{shenker, shenker1,ce,BM,shu}.
In this section we analyze the causality violation of Einstein-Weyl
gravity and obtain the new constraint imposed on the coupling
$\gamma$ of the theory.

\subsection{Tensor-type perturbation}
We rewrite the wave function as
\begin{equation}
\label{phi} \phi(t,x_3,u)=\mbox{e}^{-i\omega t+ikx_3+ik_{u}u},
\end{equation}
and take large momenta limit
$k^\mu=(\omega,k,k_u)\rightarrow\infty$. In the end, one can find
that the equation of motion (\ref{maineq}) reduces to
\begin{equation}
\label{effeq} k^{\mu}k^{\nu}g_{\mu\nu}^{\rm eff}\simeq 0,
\end{equation}
where the effective metric is given by
\begin{equation}
\dd s^2_{\rm eff} =g^{\rm eff}_{\mu\nu}\dd x^\mu\dd x^\nu
=\frac{f(u)}{L^2 z_0^2 u} \left(-\dd t^2+\frac{1}{c^2_2}\dd
x_3^2\right) +\frac{L^2}{4 u^2 f(u)}\dd u^2.
\end{equation}
Note that $c^2_2$ can be interpreted as the local speed of
helicity-$2$ graviton:
\begin{equation}
c^2_2(u)=\frac{ f^2(u) C_k(u)} {ug_T(u)}.
\end{equation}
Inserting $g_T(u)$ and $C_k(u)$ into above formula, we can expand
$c^2_2$ near the boundary $u=0$,
\begin{eqnarray}
c^2_2-1= \left(\frac{4\lambda-3}{20\lambda+3}\right)u^2
+\mathcal{O}(u^3).
\end{eqnarray}
%
%As the local speed of graviton should be smaller than $1$
% (the local speed of light), we require
%
%
\begin{figure}[h]\centering
\includegraphics[width=3in,height=3in]{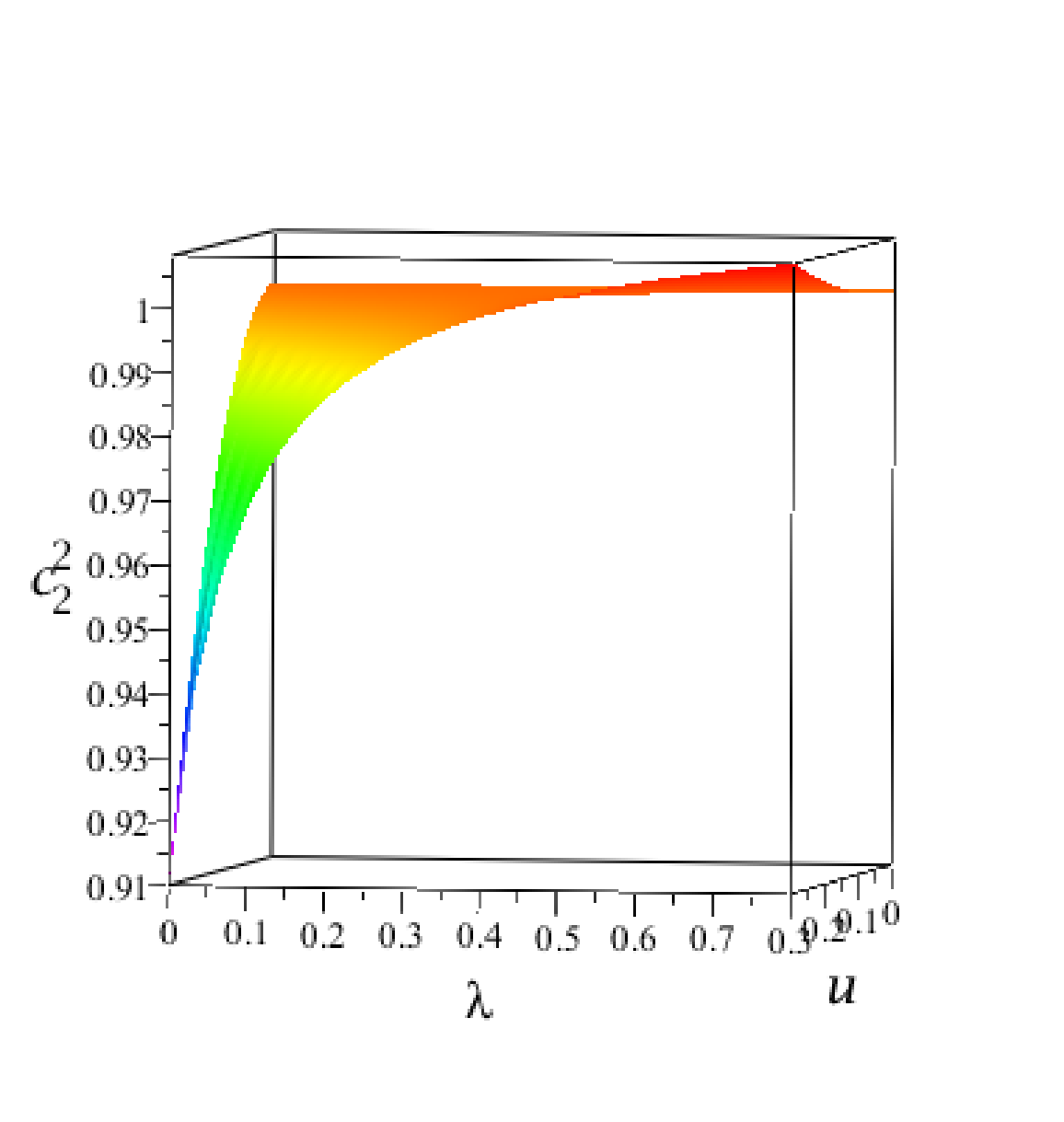}
\caption{Local speed of helicity two modes $c_2^2$ as a function of
$u$ with different $\lambda$.}\label{fig1}
\end{figure}
It was shown in \cite{shenker1} that for Gauss-Bonnet gravity the
causality violation happens when $c_2>1$, i.e., the coefficient of
$u^2$ is positive. This is because the geodesics which starts at
spatial infinity (the boundary) can bounce back to the boundary.
From AdS/CFT correspondence, it was pointed out that the
superluminal graviton propagation corresponds to the superluminal
propagation of metastable quasiparticles in the boundary CFT which
leads to the microcausality violation in the boundary CFT
\cite{shenker1}. To see it more clearly, we rewrite (\ref{maineq})
in a Schr\"{o}dinger form by introducing a new radial coordinate $y$
and $\alpha=\bar{\omega}/\bar{k}$,
\begin{eqnarray}\label{schrodinger}
-\hbar^2 \partial_{y}^2\psi+U(y)\psi=\alpha^2 \psi,\ \ \ \
\frac{dy}{du}=-\frac1{u^{1/2}f(u)},\ \ \ \ \hbar\equiv1/\bar{k},
\end{eqnarray}
where
\begin{eqnarray}
\phi=\frac1K\psi, \ \ \ \ \frac{d}{du}\ln
K\equiv\frac{g_T'(u)+16\lambda L^2
u^{3/2}f\left(u^{-1/2}f''(u)\right)'}{2g_T(u)}-\frac{(u^{1/2}f(u))'}{2u^{1/2}f(u)},
\end{eqnarray}
and
$$
U=c_2^2+\hbar^2U_1.
$$
The explicit expression for $U_1$ is too complicated and is not
necessary for our analysis below. What is relevant is that in the
limit $\hbar\rightarrow 0$ (as $k\rightarrow \infty$), only for a
tiny region $y\gtrsim -1/\bar{k}$ (or $y\gtrsim 0$) this term
becomes important. As a consequence, $c_2>1$ indicates that the
effective potential $U>1$ in the Schr\"{o}dinger problem
(\ref{schrodinger}), implying that quasinormal spectra with
$Re(\omega)/k>1$ exist in the large momenta limit. According to
AdS/CFT duality this corresponds to the presence of a pole outside
the boundary CFT light cone in the two-point function \cite{bss},
and hence lead to a violation of causality. Therefore, for NEW
gravity, the causality violation appears when
\begin{equation}\label{gamma}
 \left(\frac{4\lambda-3}{20\lambda+3}\right)>0,
\end{equation}
which indicates a constraint on $\lambda$
\begin{equation}
\lambda>\frac{3}{4}\ \ \ \mbox{or} \ \ \ \lambda<-\frac{3}{20}.
\end{equation}
However, we should be very careful to treat the results because as
we derive the results we have assumed $\lambda$ is small. In other
words, it is impossible to obtain an exact constraint on $\lambda$
from an approximated metric. What can be obtained is an approximated
constraint in our case. As a consequence, with the same
approximation, one has to expand the LHS of inequality (\ref{gamma})
to order $\mathcal{O}(\lambda^2)$,
\begin{equation}\left(\frac{4\lambda-3}{20\lambda+3}\right)= -(1-8\lambda)+\mathcal{O}(\lambda^2)>0,\end{equation}
or equivalently,
\begin{equation}
\lambda\gtrsim\frac18.
\end{equation}

Our numerical plot (\ref{fig1}) gives a visualized picture of the
relationship between local speed of helicity two modes and $u$ with
different value of coupling $\lambda$.

\subsection{Vector-type perturbation}
Similar to the above causality analysis of tensor modes, we rewrite
the wave function as
\begin{equation}
\phi(t,x_3,u)=\mbox{e}^{-i\omega t+ikx_3+ik_{u}u},
\end{equation}
and take large momenta limit
$k^\mu=(\omega,k,k_u)\rightarrow\infty$. The equation
(\ref{eom_vector}) then becomes
\begin{equation}
k_u^2+\tilde{A}_2(u)\omega^2+A_3(u)k^2\simeq 0.
\end{equation}
Then it is straightforward that the local speed of helicity one
modes is
\begin{equation}
c_1^2=\frac{\omega^2}{k^2}=-\frac{A_3(u)}{\tilde{A}_2(u)}.
\end{equation}
This leads to
\begin{equation}
c_1^2-1=-\left(\frac{3+64\lambda}{3+52\lambda}\right)u^2+\mathcal{O}(u^3).
\end{equation}
For the same reason as mentioned in tensor-type perturbations, we
cannot obtain an exact constraint on the coupling from this formula
due to the approximated black brane solution used in the paper.
However we can have an approximated constraint by assuming a small
$\lambda$
\begin{equation}
\frac{3+64\lambda}{3+52\lambda}\approx
1+4\lambda+\mathcal{O}(\lambda^2)<0,
\end{equation}
or equivalently
\begin{equation}
\lambda\lesssim-\frac14.
\end{equation}
%Our numerical result confirms this argument as shown in figure
%(\ref{fig2}).
%%
%\begin{figure}[h]\centering
%\includegraphics[width=3in,height=3in]{vector.eps}
%\caption{Local speed of helicity one modes $c_1^2$ as a function of
%$u$ with different $\gamma$ for fixed $\kappa=1$ and
%$L=1$.}\label{fig2}
%\end{figure}
%%

\section{Conclusions and Discussions}

In this paper, we have shown when small metric fluctuations around
an AdS black brane in NEW gravity are considered, the propagations
of the linearized modes impose a new bound on the coupling of the
Weyl-squared terms. Our analysis in the present paper shows that the
constraints for the scalar channel and shear channel are given,
respectively, by
\begin{eqnarray}
\rm{scalar\ channel:}&& \ \ \ \lambda\lesssim\frac18,\\
\rm{shear\ channel:}&& \ \ \ \lambda\gtrsim-\frac14.
\end{eqnarray}
Combining the above two constraints with the tachyon-free condition
(\ref{tachyonfree}) in five dimensions, one leads to the following
bounds on the Weyl-squared coupling, i.e.,
\begin{equation}
-\frac14<\lambda\lesssim \frac18,\ \ \rm{or}\ \ \
-\frac{L^2}{8\kappa}<\gamma\lesssim \frac{L^2}{16\kappa}.
\label{constraint}
\end{equation}
It is clear that the new bound imposes more stringent restrictions
than that of the tachyon-free condition, improving  predictive power
of the theory.

It is expected that the sound channel would impose another bound on
the coupling. The reason is that it was noted in \cite{hofman,BM,ce}
that for Gauss-Bonnet gravity, the upper bound of the Gauss-Bonnet
coupling comes from the tensor-type perturbations, and it is nothing
but the lower bound of the ratio of two central charges $a/c $ in
dual CFT. However, the lower bound, which comes from the scalar-type
perturbations, corresponds precisely to the upper bound of $a/c$.
Although it is not fully guaranteed to generalize this result to the
NEW gravity, one still can expect that a similar result can be
achieved. We will leave this issue for our future work.

In the present paper, we carry out computations on the gravity side
by considering small metric perturbations around an AdS black brane.
It is of great interest to compute our new bound in the context of
thermal CFTs, as what did in \cite{hofman} and \cite{ce}. This kind
of computation is important in that, on one hand, it may provide an
effective check of our results, on the other hand, it may be helpful
to our understanding about the whole picture.

We must emphasize that our calculations performed in this paper are
approximately correct. To avoid solving fourth-order differential
equations, we adopted an approximate solution twice by taking the
small $\lambda$ (or $\gamma$) limit. Since the allowing value of
$\lambda$ as given in (\ref{constraint}) is small enough, it is safe
to make this approximation (the next order contribution would be of
the order $\lambda^2\sim 10^{-2}$). In order that the result is
consistent under this approximation, one should be careful to treat
the results approximately to the same order. Although the results
are approximately correct, they are of particular importance in
understanding the causality constraint on the NEW gravity as what we
see in our result. It may provide many important clues to the future
investigation of the full analysis without any approximations.

\vspace*{10mm} \noindent
 {\large{\bf Acknowledgments}}
This work was partially supported by the NNSF key project of China
under grant No. 10935013, the NNSFC under grant Nos. 11005165 and
11175270, the SRF for ROCS under Grant No. [2009]134, the Natural
Science Foundation Projects of CQ CSTC under grant Nos. 2009BA4050
and 2009BB4084, and CQ CMEC under grant No. KJTD201016.

\vspace{1mm}

\appendix

\section{Expression for $T_{\mu\nu}^{(1)}$\label{expr_T1}}

In the expression (\ref{eom_01}), $T_{\mu\nu}^{(1)}$ is of the form
\begin{eqnarray}
\nonumber \label{eff_pert}T_{\mu\nu}^{(1)}
&=&\gamma\left\{\frac1{(D-1)(D-2)}\left(h_{\mu\nu}
{R^{(0)}}^2+2g_{\mu\nu}^{(0)}R^{(0)}R^{(1)}
-4R^{(1)}R_{\mu\nu}^{(0)}-4R^{(0)}R_{\mu\nu}^{(1)}\right)-\right.\\
&&\left.-\frac2{D-2}\left[g_{\mu\nu}^{(0)} \left(R_{\rho\sigma}
R^{\rho\sigma}\right)^{(1)}+h_{\mu\nu} R_{\rho\sigma}^{(0)}
R^{(0)\rho\sigma} -2(D-2)\left(R^\rho{}_\mu
R_{\rho\nu}\right)^{(1)}\right]\right.
\nonumber\\
&&\left.+\frac{2(D-3)}{(D-1)(D-2)}\left[g_{\mu\nu}^{(0)} (\Box
R)^{(1)}+h_{\mu\nu} \Box R^{(0)}
+(D-2)(\nabla_{\mu}\nabla_{\nu}R)^{(1)}- 2(D-1) (\Box
R_{\mu\nu})^{(1)}\right] \right.\nonumber\\ &&
\left.+\frac12g_{\mu\nu}^{(0)}\left(R_{\rho\sigma\lambda\kappa}R^{\rho\sigma\lambda\kappa}\right)^{(1)}+
\frac12h_{\mu\nu}R_{\rho\sigma\lambda\kappa}^{(0)}R^{(0)\rho\sigma\lambda\kappa}
-\frac{4(D-4)}{D-2}\left(R^{\rho\sigma}R_{\mu\rho\nu\sigma}\right)^{(1)}\right.\nonumber\\
&&\left.-2\left(R_{\mu\rho\sigma\gamma}R_{\nu}{}^{\rho\sigma\gamma}\right)^{(1)}\right\}
.
\end{eqnarray}
 where $R^{(0)}$, $R^{(0)}_{\mu\nu}$, $R_{\rho\sigma\lambda\kappa}^{(0)}$ and the covariant derivative
are defined through the background metric $g^{(0)}_{\mu\nu}$. In
(\ref{eff_pert}) the following terms are given explicitly by
\begin{eqnarray}
\left(R_{\rho\sigma} R^{\rho\sigma}\right)^{(1)}
&=&2R^{(0)\mu\nu}R_{\mu\nu}^{(1)}-2
h^{\mu\rho}R^{(0)\nu}{}_{\rho}R_{\mu\nu}^{(0)},\\
\left(R^\rho{}_\mu R_{\rho\nu}\right)^{(1)}&=&
R_{\rho\mu}^{(1)}R^{(0)\rho}{}_{\nu}+R^{(0)\rho}{}_{\mu}R^{(1)}_{\rho\nu}
-h^{\rho\sigma}R_{\sigma\mu}^{(0)}R_{\rho\nu}^{(0)},\\
(\Box R)^{(1)}&=& \Box R^{(1)}-h^{\mu\nu}\partial_\mu\partial_\nu R^{(0)}+\frac12\nabla_\mu h\cdot\nabla^\mu R^{(0)},\label{higher1}\\
(\nabla_{\mu}\nabla_{\nu}R)^{(1)}&=&\nabla_\mu\nabla_\nu
R^{(1)}-\frac12\left(\nabla_\nu h_{\mu}^\rho
+\nabla_\mu h_\nu^\rho-\nabla^\rho h_{\mu\nu}\right)\nabla_\rho R^{(0)},\\
\nonumber(\Box R_{\mu\nu})^{(1)}&=&\frac12\Box
R_{\mu\nu}^{(1)}+\frac12\left(\nabla^\rho\nabla^\sigma
h_{\rho\mu}-\Box h_{\mu}^{\sigma} -\nabla^\rho\nabla_\mu
h_\rho^\sigma\right)R_{\nu\sigma}^{(0)}\\
\nonumber&&+\left(\nabla^\sigma h_{\rho\mu}-\nabla_\rho
h_{\mu}^{\sigma} -\nabla_\mu h_\rho^\sigma\right)\nabla^\rho
R_{\nu\sigma}^{(0)}-\frac12\nabla_\rho h^{\rho\sigma}\nabla_\sigma R_{\mu\nu}^{(0)}\\
&&+\frac14\nabla^\rho h\nabla_\rho R_{\mu\nu}^{(0)}-\frac12
h^{\rho\sigma}\nabla_\rho\nabla_\sigma R_{\mu\nu}^{(0)}+(\mu
\leftrightarrow \nu),\label{higher2}\\
\left(R_{\rho\sigma\lambda\kappa}R^{\rho\sigma\lambda\kappa}\right)^{(1)}&=&2R^{(0)\mu\nu\rho\sigma}R_{\mu\nu\rho\sigma}^{(1)}
-4h^{\mu}_{\kappa}R_{\mu\nu\rho\sigma}^{(0)}R^{(0)\kappa\nu\rho\sigma},\\
\left(R^{\rho\sigma}R_{\mu\rho\nu\sigma}\right)^{(1)}&=&
R^{(0)\rho\sigma}R_{\mu\rho\nu\sigma}^{(1)}+R_{\rho\sigma}^{(1)}R^{(0)}{}_{\mu}{}^{\rho}{}_{\nu}{}^{\sigma}
-h^{\rho\kappa}R^{(0)}{}_{\kappa}{}^{\sigma}\left(R^{(0)}_{\mu\rho\nu\sigma}+R^{(0)}_{\mu\sigma\nu\rho}\right),\\
\nonumber\left(R_{\mu\rho\sigma\gamma}R_{\nu}{}^{\rho\sigma\gamma}\right)^{(1)}&=&R^{(1)}_{\mu\rho\sigma\gamma}R^{(0)}{}_{\nu}{}^{\rho\sigma\gamma}+
R^{(1)}_{\nu\rho\sigma\gamma}R^{(0)}{}_{\mu}{}^{\rho\sigma\gamma}-\\
&&-R^{(0)}_{\mu\rho\sigma\gamma}\left(h^\rho_\kappa
R^{(0)}{}_{\nu}{}^{\kappa\sigma\gamma}+h^\sigma_\kappa
R^{(0)}{}_{\nu}{}^{\rho\kappa\gamma} +h^\gamma_\kappa
R^{(0)}{}_{\nu}{}^{\rho\sigma\kappa}\right),
\end{eqnarray}
where
$$
R^{(1)}_{\kappa\mu\sigma\nu}=h_{\rho\kappa}R^{(0)\rho}{}_{\mu\sigma\nu}+\frac12\left[\nabla_\nu\nabla_\kappa
h_{\mu\sigma}-\nabla_\nu\nabla_\mu
h_{\kappa\sigma}-\nabla_\sigma\nabla_\kappa
h_{\mu\nu}+\nabla_\sigma\nabla_\mu
h_{\kappa\nu}+2\nabla_{[\sigma}\nabla_{\nu]} h_{\kappa\mu}\right].
$$

\section{Expressions for $A_i(u)$ and $B_i(u)$\label{expr for A}}
Expressions for $A_i(u)\ (i=2,3,5)$ are given respectively, by
\begin{eqnarray}
A_2(u)&=&\frac{8\lambda z_0^2L^6\left(3-4f(u)-u^2f''(u)+3u
f'(u)\right)}{u f(u)^2g_V(u)},\\
A_3(u)&=&\frac{3z_0^2L^6\left(1+\frac{\lambda}6(68u^2f''(u)-218u f'(u)+296f(u)-192)\right)}{4 u f(u) g_V(u)},\\
A_5(u)&=&\frac{3z_0^2L^6\left(1+\frac{\lambda}6(4u^2f''(u)-26u
f'(u)+40f(u))\right)}{4 u f(u) g_V(u)},
\end{eqnarray}
where $g_V(u)$ is
\begin{equation}
g_V(u)=-3L^2-\frac{\lambda L^2}{2}\left(36u^2f''(u)-198u
f'(u)+296f(u)-192\right).\label{gv}
\end{equation}
Expressions for $B_i(u) (i=1,2)$ are given, respectively, by
\begin{eqnarray}
B_1(u)&=&-\frac{2 \left(14u^2f''(u)+3u
f'(u)-20f(u)\right)}{3L^2},\\
B_2(u)&=&-\frac{2 \left(-2u^2f''(u)+13u f'(u)-20f(u)\right)}{3L^2}.
\end{eqnarray}

\end{document}